# Antisymmetry in hydrogenic HH$^+$ and antihydrogenic $\underline{H}$H$^+$ cations: classical and ab initio quantum mechanical calculations

G. Van Hooydonk, Ghent University, Faculty of Sciences, Krijgslaan 281, B-9000 Ghent (Belgium)

Abstract. Ab initio quantum chemistry reveals how the charge-antisymmetric antihydrogenic $\underline{H}$H$^+$ state, deriving from conventional positional antisymmetry of a 3-unit charge system, is hidden in the PEC (potential energy curve) of the molecular hydrogen cation. Only with internal charge inversion and Coulomb's law, the anti-symmetry generated by positional coordinates can be understood but this solution was persistently overlooked. Stabilizing matter-antimatter interactions not only exist in nature, they are an essential element, even in wave mechanics, for explaining stable composite matter with a Coulomb model.

**Introduction**

Stating that matter-antimatter systems must not annihilate [1] contradicts the so-called matter-antimatter asymmetry of the Universe [2]. Recent claims on the production of neutral antimatter (*artificial antihydrogen* $\underline{H}$) [3] boosted work on matter-antimatter interactions like H-$\underline{H}$ to calculate their PECs (potential energy curves) [4]. Yet, these H$\underline{H}$ PECs are contradictory on the issue of cusp formation, despite the highly sophisticated quantum mechanical calculations used [1,5]. Anti-symmetry is important for the stability of quantum systems but it can be due to *positional coordinates* (parity **P**), *spin symmetries* (**S**), *charge symmetries* (**C**)… With intra-atomic charge-inversion, one expects stable *charge-symmetric* HH to transform in *unstable and exotic charge-antisymmetric* $\underline{H}$H [1]. We now verify how a trivial Coulomb anti-symmetry scheme is hidden in the framework of wave mechanics for the molecular hydrogen cation (charged 3-particle Coulomb system). Historically, the wave mechanical analysis for this *charged two center* system, as well as for *neutral* 3-unit charge *one center* Coulomb system He, proved decisive for the development of quantum mechanics for many particle systems [6]. We focus on the simplest bond imaginable, *charge-symmetric hydrogenic cation* HH$^+$, extensively studied theoretically and experimentally [6] and on *charge-anti-symmetric or antihydrogenic cation* $\underline{H}$H$^+$, unknown in the early days of wave mechanics.

**Coulomb-based antisymmetry in the molecular hydrogenic cation**

Anti-symmetry by intra-atomic charge inversion invariantly leads to a parity operator (algebra) within the 2-center 3-particle Hamiltonian for the molecular hydrogen cation

$$\mathbf{H}_\pm = \mathbf{H}_0 \pm \Delta\mathbf{H} \qquad (1a)$$

exactly as for 2-center 4-unit charge system H$_2$ [1]. The result for the cation is simpler, since there are only 4 terms, instead of 10 for H$_2$ [1], and only 2 Coulomb terms are perturbing, since

$$\pm\Delta\mathbf{H} = \pm(-e^2/r_B + e^2/r_{AB}) \qquad (1b)$$

The + solution applies for *charge-symmetric hydrogenic* HH$^+$, the negative for *charge-anti-symmetric antihydrogenic* $\underline{H}$H$^+$. Lepton-nucleon separation $r_A$ appears in $\mathbf{H}_0$ for neutral atom H or $\underline{H}$ (Coulomb potential energy $-e^2/r_A$), whereas lepton-nucleon separation $r_B$ appears in a perturbing term, just like nucleon-nucleon separation $r_{AB}$ in both Hp$^+$ and $\underline{H}$p$^+$. The proton is at the second center because of the total cation charge.

Many will find it difficult to accept that an *exotic antihydrogenic* state like $\underline{H}$H$^+$ is needed to account for the stability of the observed molecular hydrogen cation [1]. Hence, (1a) must also be tested with wave mechanics to see if it makes sense. It is obvious from (1b) that the *observed* PEC of the cation must be checked for a the presence of a cusp, pointing to a switch from *repulsive charge-symmetric* $+e^2/r_{AB}$ for $HH^+$ to *attractive charge-anti-symmetric* $-e^2/r_{AB}$ for $\underline{H}H^+$, in order to validate (1b) and the many drastic consequences it may have [1].

**Generic Coulomb cusp for matter-antimatter antisymmetry**

Finding a signature for a (reversible) but discrete hydrogen-antihydrogen transition of type
$$+e^2/r_{AB} \text{ (for } HH^+\text{)} \leftrightarrow -e^2/r_{AB} \text{ (for } \underline{H}H^+\text{)} \quad (1c)$$
between two mutually exclusive states for interacting nucleons, *anywhere in the observed PEC of the molecular hydrogen cation*, is essential. This makes (1c) of crucial importance to validate [1]. It would be nice if we could predict exactly where transition (1c) must occur and how this could be understood with first principles, say the classical Coulomb law.

To understand the formation of a molecular hydrogen cation, two different asymptotes are required. Their difference is an energy gap of $-½$ Hartree, i.e. $\mathbf{H}_0 = -½e^2/r_A = -½e^2/a_0$ in (1a). With separation R between the unit charges in au, the scaled variable is $D = R/a_0$. Obviously, Coulomb attraction $\sim -1/D$ sets in at asymptote 0 for infinite separation, whereas repulsion $\sim +1/D$ must set in at asymptote $-1/2$ Hartree, giving interaction
$$W_+(D) = -½ + 1/D = -½(1-2/D) \quad (1d)$$
which approaches the zero asymptote from below. By Coulomb symmetry (parity $\pm 1$), the associated attractive interaction approaches the same zero asymptote from above or
$$W_-(D) = -(-½(1-2/D)) = ½(1-2/D). \quad (1e)$$
Similarly, the scaled classical Coulomb interactions viewed from the zero asymptote now becomes repulsive $+2/D$ for hydrogenic $HH^+$ and attractive $-2/D$ for antihydrogenic $\underline{H}H^+$. The general naïve Coulomb picture for (1a) is completely described with PECs obeying
$$W_\pm(D) = \pm(1-2/D) \quad (1f)$$
the basis of our discussion of (1b) and to be confronted with wave mechanics (see below).

The four classical PECs of type W(D), e.g. $\pm 2/D$ and $\pm(1-2/D)$ for unit asymptotes $\pm 1$ instead of $\pm ½$ as in (1d) and (1e), are depicted in Fig. 1a with variable D and in Fig. 1b with $1/D$ [7]. This is the rather trivial effect of the Coulomb law for 2 unit charge systems. In either scheme, the zero asymptote is reproduced exactly by summing conjugated Coulomb PECs like $+2/D - 2/D = 0$. In this sense, one would conclude that matter would annihilate with anti-matter although, strictly spoken, only the pair-wise conjugated point charges, *deriving from an imposed Coulomb model to describe the particles*, are annihilated (see below).

Yet there are differences between representations with and without asymptotes $+1$ and $-1$, connected by the parity operator $\mathbf{P}$, deriving from interacting charges or from $\mathbf{C}$:
(i) Unlike the model with the zero asymptote, only the model with 2 asymptotes $\pm 1$ is capable of predicting a cusp for the system. By symmetry, this Coulomb cusp is theoretically fixed exactly at 2D, i.e. $R=2a_0$, *which is also exactly the observed equilibrium separation of the molecular hydrogen cation* [6].
(ii) And if this generically predicted cusp for the cation makes sense, the two asymptotes must make sense also. This means in concreto that the quantum state with asymptote $+1$ applies to *antihydrogenic* $\underline{H}H^+$, whereas asymptote $-1$ is valid for the quantum state with *hydrogenic* $HH^+$. But this also means that *both states (or asymptotes) $\underline{H}H^+$ and $HH^+$ are required to understand the naturally observed molecular hydrogen cation, e.g. its Coulomb cusp*. These bold but inescapable conclusions from (1a)-(1c) must be confronted with wave mechanics (see below).

**Cusps with generalized Coulomb interactions**

Changing from potential energies $\pm 1/D$ to forces $\pm 1/D^2$ (or to any conjugated pair of another type[1] of interaction $\pm 1/D^n$, even enforced or attenuated symmetrically…) will never alter conceptually the PEC behavior in Fig. 1 for a stable unit system, since the generic cusp at D=2 is always conserved by symmetry. For instance, we could easily mimic a slowly attenuated interaction using a continuously varying function like $e^{-D}$ (see further below).

---
[1] In this sense, an interaction varying as $D^0$, $D^1$, $D^2$…(like Hooke's law), cannot be excluded a priori [7].

*Whatever the degree of attenuation or whatever the value of n in $1/D^n$, the position of the Coulomb cusp remains invariantly at D=2. This mechanism can become an important trompe-l'oeil effect in the interpretation of PECs for stable systems [7b]. In particular, a generic cusp at D=2 in any PEC can always be interpreted as deriving from Coulomb charge-anti-symmetry and its two conjugated asymptotes or Coulomb quantum states ±1. As soon as the pure Coulomb law is not obeyed exactly, the zero asymptote is no longer reproduced by adding attractive and repulsive branches and, instead, a PEC of type W(D) will have to appear around the zero asymptote. Only an asymmetry in the repulsive and attractive branches can, eventually, also distort the position of the Coulomb cusp.*

With Fig. 1, the anti-symmetry problem due to intra-atomic charge inversion obeying Coulomb's law is quantitatively well defined classically. We must analyze in detail the observed and calculated PECs for unit charge Coulomb systems, *charge-conjugated cations* $HH^+$ and $\underline{H}H^+$, verify if a generic cusp at D=2 like in Fig. 1 can be detected and, finally, how this is interpreted with wave mechanics.

**Coulomb splitting with 2 quantum states ±1: ionic or covalent bonding models**

Coulomb recipe (1f) is one of the oldest examples of 2 charge-conjugated quantum states +1 and –1 possible for a neutral system. In essence, it is a recipe for describing a neutral system at the zero asymptote (say axis x) with the help of two new asymptotes (virtual or real) at ±1, *obtained after symmetrically unfolding the origin along the y-axis if the system is on the x-axis*. This creates an extra symmetrical biaxial reference frame to describe the system.

The oldest application in chemistry is ionic bonding between cation $A^+$ and anion $B^-$, interacting by means of Coulomb's law $-1/r_{AB}$. As discussed in full elsewhere [7], the complete PEC for this interaction has an attractive branch for a charge-antisymmetric system (ion-anti-ion system) following $-1/r_{AB}$, whereas its repulsive branch for the charge-symmetrical system follows $+1/r_{AB}$ with a cusp at $r_{AB}=2r_0$, the equilibrium separation. This is the basis of an ionic approximation to chemical bonding. In first order, this ionic PEC can be generated very easily with the Kratzer potential, even for covalent molecules [1,7].

*In contrast to ionic bonding schemes, covalent bonding, proposed by Heitler and London [8], explains bonding directly from the zero asymptote, i.e. seemingly without any assistance of the Coulomb asymptotes ±1, typical for ionic bonding and, finally, also seemingly without the recipe provided by (1f) and the PECs as described in Fig. 1. This is why ionic bonding models were immediately abandoned and superseded with the Heitler-London (HL) model, published in 1927, although we argued that this may well have been an unjust decision* [1].

*This controversy on chemical bonding (covalent or ionic) can now be made very concrete: does wave mechanical HL theory rely or not on the Coulomb PECs with asymptotes or quantum states ±1, respectively for the atom-atom and the atom-anti-atom states as depicted in Fig. 1?*

In fact, we will show below that, despite all appearances, even wave mechanical HL-theory is based upon this very same classical Coulomb recipe (1f) and its old fashioned ionic but generic PECs as depicted in Fig. 1. To prove this analytically, we only need a simple transparent example of quantum mechanical Heitler-London bonding, which is provided with the one lepton two center molecular hydrogen cation.

**Ab initio wave mechanical calculation of the PEC for the hydrogen molecular cation**

To retrace the essentials of the PECs in Fig. 1 in a wave mechanical framework, we need a standard calculation with straightforward interpretation of all terms evaluated. Fortunately, Pauling and Wilson [6] proposed a very transparent wave mechanical analysis for $HH^+$, similar to that used by Heitler and London for $H_2$ [8] and which we will therefore use here in the greatest detail. This PW-model [6] analysis starts with the typical secular equation

$$\begin{vmatrix} H_{AA} - W & H_{AB} - SW \\ H_{BA} - SW & H_{BB} - W \end{vmatrix} = 0 \quad (2a)$$

wherein
$$H_{AA} = \int u_{1s(A)} H u_{1s(A)} d\tau \quad (2b)$$
$$H_{BB} = \int u_{1s(B)} H u_{1s(B)} d\tau \quad (2c)$$
$$H_{AB} = \int u_{1s(A)} H u_{1s(B)} d\tau \quad (2d)$$
$$H_{BA} = \int u_{1s(B)} H u_{1s(A)} d\tau \quad (2e)$$
$$S = \int u_{1s(A)} u_{1s(B)} d\tau \quad (2f)$$

S represents the lack of orthogonality of hydrogenic atomic functions $u_{1s(A)}$ and $u_{1s(B)}$, referred to above (Fig. 1) as the appearance of a biaxial reference system. Because of the pair-wise positional (geometrical) equivalence, $H_{AA} = H_{BB}$ and $H_{AB} = H_{BA}$ hold [6]. Standard solutions for (2a) are either symmetric (S) or anti-symmetric (A), i.e.

$$W_S = (H_{AA} + H_{AB})/(1+S) \quad (3a)$$
$$W_A = (H_{AA} - H_{AB})/(1-S) \quad (3b)$$

for, respectively, symmetric (S) and anti-symmetric (A) wave functions

$$\psi_S = (u_{1s(A)} + u_{1s(B)})/(2+2S)^{1/2} \quad (4a)$$
$$\psi_A = (u_{1s(A)} - u_{1s(B)})/(2-2S)^{1/2} \quad (4b)$$

due to the fact that only positional coordinates are used.
Solutions (3) derive from the quadratic in (2a), leading to

$$(H_{AA}-W)^2 - (H_{AB}-SW)^2 = 0$$
$$W(1-S^2) - 2W(H_{AA}-SH_{AB}) + H_{AA}^2 - H_{AB}^2 = 0$$
$$W = (1/(1-S^2))[H_{AA} - SH_{AB} \pm (H_{AB} - SH_{AA})] \quad (4c)$$

Since there is only one lepton, spin symmetries must not be considered to secure anti-symmetry of the total wave function for a 2 center 2-lepton system like $H_2$ [1,6]. We must only verify if there is a correspondence between anti-symmetry generated by positional coordinates or by charge inversion[2] (when *charge-symmetrical* state H $e^-p^+$ goes over in *charge-anti-symmetrical* $\underline{H}$ $e^+p^-$, as in (1c) and as argued in [1]).
Following [6], we introduce eigenvalue $W_H$, given by

$$-(h^2/8\pi^2 m_0)\nabla^2 u_{1s(A)} - (e^2/r_A) u_{1s(A)} = W_H u_{1s(A)}$$

reminding this relation should, in a reasonable first approximation, apply for both H and $\underline{H}$ [1].
Due to the relative simplicity of the cation, all integrals are easily calculated analytically [6]:

$$H_{AA} = \int u_{1s(A)} (W_H - e^2/r_B + e^2/r_{AB}) u_{1s(A)} d\tau$$
$$= W_H + e^2/a_0 D + J \quad (5a)$$

wherein
$$J = \int u_{1s(A)} (-e^2/r_B) u_{1s(A)} d\tau = (e^2/a_0)(-1/D + e^{-2D}(1+1/D)) \quad (5b)$$

with a scaled notation for the nucleon-nucleon separation
$$D = r_{AB}/a_0 \quad (6)$$

already introduced above for the construction of the Coulomb schemes in Fig. 1.
We notice that, *without further differentiation by scaling*, proton-proton repulsion $+e^2/r_{AB}$ would be exactly canceled by the attractive term $-e^2/r_{AB}$ in J (but see further below).
Similarly

$$H_{BA} = \int u_{1s(B)} (W_H - e^2/r_B + e^2/r_{AB}) u_{1s(A)} d\tau$$
$$= SW_H + K + Se^2/a_0 D \quad (7a)$$

wherein
$$K = \int u_{1s(B)} (-e^2/r_B) u_{1s(A)} d\tau = (-e^2/a_0) e^{-D}(1+D) \quad (7b)$$

and the lack of orthogonality (introduction of a biaxial reference system) obeys
$$S = e^{-D}(1+D+D^2/3) \quad (8)$$

---

[2] It is important to realize that in the 1930s [6], the concept of antiparticles in the newly developed theory of the $H_2$ chemical bond by Heitler and London in 1927 [8] was much too new to be incorporated as a possibility for bonding, as remarked in [1].

With all integrals available, standard solutions (3) follow immediately. Exponential $e^{-D}$ in (8) could eventually attenuate Coulomb interactions without affecting the position of the cusp at D= 2 (see above).

In the PW scheme, most if not all terms for the nucleonic interaction are repulsive $+e^2/r_{AB}$. This will make it difficult, if not impossible, to find evidence for the reality of cusp (1c) at D=2 and for the antihydrogenic asymptote explicitly needed for attractive $-e^2/r_{AB}$ to make sense.

In the hypothesis that the hydrogenic molecular cation be denoted as $HH^+$ (or $H_2^+$), (5)-(8) lead to the well-known symmetric (S) and anti-symmetric (A) solutions for (3)

$$W_S = W_H + e^2/a_0D + (J+K)/(1+S) \qquad (9a)$$
$$W_A = W_H + e^2/a_0D + (J-K)/(1-S) \qquad (9b)$$

as illustrated in Fig. 42-2 of [6]. Splitting of bonding and anti-bonding states is caused by virtue of the 2-center *exchange or resonance integral* ±K (also called overlap), although K itself is a continuous function in the range $0<D\leq+\infty$. If so, K itself cannot be responsible for a cusp. Also overlap S is a continuous function in the same range, devoid of any cusp.

A conventional W, D representation for S and A solutions (9) is given in Fig. 2, which illustrates the achievements of wave mechanics for this problem [6]. The solutions, *unknown in classical physics*, seem to derive solely from quantum theory. Pauling and Wilson repeatedly argue that Heitler and London used a similar procedure [8] to explain $H_2$ bonding with wave mechanics, a solution we revisited recently [1].

The cusp problem (1c) can only be solved by adding/subtracting some of the primary integrals above (1/D, J, K, S…). The PECs generated by all integrals are given in Fig. 3 to verify that all are rather continuous without any cusp.

To illustrate also the behavior of functions $H_{AA}$ (5a) and $H_{AB}$ (7a), we give their PECs in Fig. 4. It is evident that by adding them a cusp will be formed in the end but $H_{AA}$ is continuously repulsive whereas $H_{AB}$ is attractive at long range but becomes repulsive at shorter range, indicating that cusp formation in wave mechanics relies solely on $H_{AB}$.

After some mathematical manipulation (scaling and shifting it to asymptote zero instead of $W_H$), it is easily shown that $W_S$ in (9a) leads to a rather cryptic end result

$$W'_S = +2\{S/D + e^{-D}[e^{-D}(1+1/D)-(1+D)]\}/(1+S) \qquad (9c)$$

which can hardly be recognized with classical physics. This PEC is given in Fig. 2. It does not bear any resemblance with the elementary even trivial Coulomb functions (1f), given in Fig. 1. Equations like (9c) were considered as *a triumph for wave mechanics* in the case of $H_2$ [6], as they seemed to be completely out of reach for classical physics. This, however, happens to be a misjudgment [1], as we show below.

Our proof relies mainly on scaling. If, for instance, we would have scaled (9a) and (9b) with absolute[3] $|W_H| = |e^2/2a_0|$, the mathematically correct scaled versions are

$$W_S/|W_H| = -1 + 2/D + 2(J+K)/(1+S) \qquad (9d)$$
$$W_A/|W_H| = -1 + 2/D + 2(J-K)/(1-S) \qquad (9e)$$

suggesting indeed, just like everybody would have expected, that wave mechanics exclusively uses the repulsive interaction $+e^2/r_{AB}$, typical for the $HH^+$ asymptote for the molecular hydrogen cation and, therefore, would have excluded asymptote +1 for $\underline{H}H^+$ with its attractive Coulomb interaction $-e^2/r_{AB}$ from its calculations. If this were really true, wave mechanics could never have made full advantage of the generic cusp condition deriving from the Coulomb scheme above. As a matter of fact, the cusp condition $-1+2/D=0$ applied to (9d) and (9e) would only generate a crossing point with the zero asymptote, without any further symmetry implication and, especially, without having to refer to the Coulomb symmetry for 2 asymptotes or 2 quantum states ±1 (see above).

---

[3] It is easily verified that scaling with the eigenvalue $W_H = -e^2/2a_0$ would invert the character of all interactions involved. Attractive would become repulsive, and repulsive attractive. Even with this odd assumption, the more essential two asymptote problem ±1 can never be never avoided.

## D=2 Coulomb cusp for a hydrogen-antihydrogen transition in ab initio wave mechanics: analytical method

The real problem is therefore to retrace exactly the nature of the wave mechanical cusp in Fig. 2 for bound state (9a). In [1], we collected all repulsive and attractive terms in the Hamiltonian to arrive at a molecular Kratzer potential. We can now refine the same procedure more rigorously with all the terms and their values in (9a), obtained with the PW *ab initio* wave mechanical calculation. The true solution for the Coulomb cusp requires only three very elementary mathematical but physically very essential steps, *which we could not yet retrace in the literature.*
(i) Let us first reposition $W_S$ in (9a) at the zero asymptote by adding $-½W_H$, giving $W^0_S$ or
$$W^0_S = W_S - ½W_H \qquad (10a)$$
(ii) Let us now scale like for (9d) and (9e). Since $W_H = -½e^2/a_0$, scaling (10a) with $|W_H| = +½e^2/a_0$ gives a different result as in (9d). This is $W'_S = W^0_S/(+e^2/2a_0)$ or
$$W'_S = -1 + 2/D + 2(J' + K')/(1+S) + 1 \qquad (10b)$$
in which $J' = J/(e^2/2a_0)$ and $K' = K/(e^2/2a_0)$. Unlike (9a) and (9d), function (10b) starts from the zero asymptote at large D. For some strange reason to become clear below, we did not yet decide to replace +1-1 in (10b) by 0.
(iii) For the last step, we refer to (5b), use the identity $J' = -2/D + J''$ with $J'' = +2e^{-2D}(1+1/D)$, respect the sign of K and obtain instead of (9a) its mathematical equivalent
$$W'_S = \{[-1+2/D]+2e^{-2D}(1+1/D)/(1+S)\}_r$$
$$+ \{[1-(2/D)/(1+S)] + 2e^{-D}(1+D)/(1+S)]\}_a \qquad (10c)$$
where we have managed to collect *all repulsive terms* belonging to asymptote or quantum state $-1$ $HH^+$ (subscript r) and *all attractive terms* for asymptote or quantum state $+1$ $\underline{H}H^+$ (subscript a). Square brackets give the terms corresponding with the classical Coulomb functions $(-1+2/D)_r$ and $(1-2/D)_a$ used for PECs (1f) in Fig. 1.
We immediately verify that, leaving out the asymptotes +1 an –1, the two functions start off as $±2/D$, exactly as in Fig. 1 for the classical Coulomb recipe.
This means that, like in (9d), the main and first repulsive term between square brackets is analytically exact the repulsive term in the asymptote directed Coulomb scheme in Fig. 1 for asymptote or quantum state $-1$, i.e. $HH^+$. However, the new result is that the main attractive term between the second pair of square brackets, *necessarily belonging to mutually exclusive Coulomb quantum state +1 or $\underline{H}H^+$*, only differs through the appearance of factor $1/(1+S)$, which reduces to 1 for small S (at large D). For small S, this term is equal to $+1-2/D$ as in (1f). Looking at the first principles involved, we cannot but conclude, analytically, that at long range, where the perturbation of atomic H or $\underline{H}$ by a proton sets in, the classical Coulomb approach (1f) *with its inescapable antihydrogenic asymptote*, is hardly discernable from the wave-mechanical approach (10c). As explained above, only in the more complex format (10c), taken directly from wave mechanics *although it is hidden therein*, it is possible to position a really generic Coulomb cusp at $R=2a_0$ needed to explain the stability of the molecular hydrogen cation.
The generic Coulomb cusp condition of Fig. 1 would have remained invisible if we had directly (be it correctly) simplified (9a) to analytical form (9c). But by rushing into (9c) by overlooking intermediary (10c) from which it derives, all essential information pointing towards Coulomb scheme Fig. 1, and in particular this important cusp formation mechanism would have been lost. Then, denying that this strictly wave mechanical scheme (10c) essentially derives from a seemingly naive Coulomb antisymmetry would be unfair. This negation, by hiding the first principles Coulomb model, would have led us indeed to the premature and unjust conclusion that only wave mechanics can cope with the essentials of chemical bonding. Failing to have seen this at an earlier stage with a simple exercise as done here and whereby the hydrogen-antihydrogen interaction must play a fundamental role, is exactly what led us the revisit the Heitler-London model for $H_2$ [1].

**Results and discussion**

The PEC resulting from (10c), identical with PW-relation (9a), is given in Fig. 5a for variable D and in Fig. 5b for 1/D. In this form, it can directly and quantitatively be confronted with the classical Coulomb scheme given in Fig. 1.

We easily see in Fig. 5a that the repulsive contributions in the naïve Coulomb scheme and in the ab initio PW scheme are almost identical in the complete D-range. The attractive branch is slightly more distorted in order to produce the PEC given by (9c). These details are more visible in Fig. 5b.

But the result is particularly interesting for the cusp formation mechanism. With a Coulomb scheme, this is almost trivial but would be mathematically impossible with the standard wave mechanical procedure *if one would adhere only to end result* (9c). The Coulomb cusp is, by definition, exactly at $R=2a_0=1,06$ Å, the observed value [6], whereas the PW-result is much larger 1,32 Å [6]. *In Coulomb terms, this means that the PW-approach fails to respect the strict Coulomb cusp condition by wrongly estimating the deviations from pure Coulomb behavior, shown in Fig 5*. And looking at (10c), it is easily verified that this error can, analytically, *only reside in the wave functions used*, as argued in [1].

The PW-result for the well depth for the bound state of the molecular hydrogen cation is 1,77 eV, whereas experiment gives 2,78 eV [6]. Although we predicted the cusp correctly at a very early stage (see Fig. 1), we admit we cannot yet give an alternative classical solution for the well depth. Yet, we see that the analytical form of the final (9c) PW-solution is completely dependent on the analytical form of their wave functions (hydrogenic STOs) [5]. However, it can be anticipated with *classical physics* that, around the cusp, the simple Coulomb law will have to be adapted anyhow. In fact, in the neighborhood of the Coulomb cusp at D=2, it can be expected *with classical physics indeed* that the (circular) motion of the lepton will interfere with the Coulomb force. It is evident that kinetic energies are always repulsive and of type $+b/r^2$ [1].

To quantify these classical views, we can easily give 2 phenomenological examples, inspired by the Kratzer and Morse potentials, *since both potentials are also centered at the Coulomb cusp* [7b], predicted by the matter-antimatter scheme in Fig. 1.

Using the observed well depth of 2,78 eV [6], we looked for *slowly varying* cusp-respecting perturbing functions and found *good behaving functions* $0,15S[(1\pm2/D)^2-1]$ ( *exponentially attenuated* Kratzer-type) and $4,5e^{-D}(e^{-D}\pm0,3)$ (*exponential* Morse-type). In each case, the + sign is correlated the repulsive or anti-bonding state (without cusp), the – sign with the bound or attractive state (with cusp), as prescribed in [1]. The 2 *phenomenological* PECs for bonding and anti-bonding states of Kratzer and Morse type are all depicted in Fig. 6, together with the PW functions (9a)-(9b) to accentuate the differences possible at the level of PEC construction.

It is obvious from Fig. 6 that other asymmetries for or deviations from the original Coulomb law are, at least, possible and even plausible. We will not discuss here in full the differences between Morse and Kratzer functions, since this was done extensively elsewhere [7b]. In addition, wave mechanics is still confronted with the problem of the analytical differences between STOs and GTOs [5], which will greatly affect the final outcome of PW-functions like (9c). Also this problem cannot be discussed here in further detail [5].

The analytical development for state $W_A$ runs completely parallel to that given for (10c). Only here, the attraction is greatly enhanced, even more than for state $W_S$, but the enhancement for the repulsive terms turns out to be even greater, which makes this a repulsive state without extreme. These extremely interesting details on wave mechanical procedures, inspired by Coulomb behavior in Fig. 1, were never mentioned in the literature: this is why, conventionally, chemical bonding is still regarded as a one asymptote problem which can only be left with a repulsive force $+e^2/r_{AB}$. Our results prove why this restricted view must not be true after all.

Moreover, looking for deviations of pure Coulomb behavior is not an exclusivity of wave mechanics. It is known for long from classical physics/chemistry that the original formulation of the Coulomb potential 1/R or force $1/R^2$ can be put in doubt, since a series with a number terms like $A/R^n$ is very well possible [7a], as is evident from the many molecular potentials proposed in the literature throughout the years [7] (see also the Kratzer function above). Realizing this, there is a great resemblance between classical physics and wave mechanics, typified by Fig. 1 and 5. Both are trying to find out more about deviations from Coulomb's law. In classical physics, this is done *by trial and error* to find a universal molecular function or *universal equation of state* (UEOS) using spectral data and the PEC they generate [7]. In wave mechanics, this is done *by trial and error* by looking for the best wave functions possible using the same spectral data and the same PEC. It is now certain by all means that standard *ab initio* wave mechanics *almost secretly* but nevertheless formally and analytically admits the reality of and the need for the *antihydrogenic* $\underline{H}H^+$ asymptote as shown by the resemblance of Fig. 5 with Fig. 1. We easily verify that the wave mechanical solution tries to quantify the small deviations from pure Coulomb behavior with its 2 asymptotes as in Fig. 1, which is a conclusion never made before, to the best of my knowledge. This result makes it the more difficult to understand why the theoretical physics and chemistry establishment are so strongly opposed to bonding schemes using the crucial matter-antimatter asymptote, apparent from the first Hamiltonian symmetry in (1a), suggested in [1].

Finally, the Coulomb interaction between charged species plays an important role *only in the ionic approximation to chemical bonding* [1]. Here, an anion interacts with a cation by means of Coulomb – 1/R attraction, which, in essence, is *an annihilative charge-conjugated ion-anti-ion Coulomb interaction* in the original Dirac sense [1].

The confrontation of Fig. 5 with Fig. 1 leads to the same conclusion as Coulson's [9] that '*it has been laughingly said that quantum mechanics weighs the captain of a ship by weighing the ship when he is and is not on board*'. In this terminology, the *weight of the ship* corresponds with the classical Coulomb anti-symmetry scheme in Fig. 1. Its slight distortions, *seemingly hidden* in its wave mechanical version in Fig. 5, would correspond with the *weight of the captain*.

The zero asymptote approximation used in the complex apparatus of quantum chemistry as we know it today is devoted almost exclusively *to the relatively small deviations* in Fig. 5 *from the original and generic Coulomb recipe in Fig. 1 with its matter-matter and matter-antimatter asymptotes and the transitions between them*. Zooming in only on these otherwise interesting deviations of pure Coulomb behavior *at large* as in Fig. 5, fails to recognize the importance of the global Coulomb $\pm 1/R$ scheme itself for chemical bonding. This is even more apparent when looking at the shape of the Coulomb PECs in Fig. 1 and *at the pure Coulomb –1/R interaction, typical for centuries old ionic or ion-anti-ion approximations to chemical bonding* [1]. Exploiting consistently this strange similarity between pure *ion-anti-ion* and generic Coulomb PECs, leads straightforwardly to other approaches to chemical bonding [1,7a] with essentially the same atom-anti-atom asymptotes and interactions in (1f). Despite appearances, matter-antimatter asymptotes are used *identically but secretly* in the covalent HL wave mechanical approximation [1]. Ionic and covalent bonding models rely on the very same Coulomb symmetry principles (1f) and on the very same matter-matter and matter-antimatter asymptotes or quantum states ±1, which is exactly as we argued in [1].

**Conclusion**

We proved explicitly that generic algebraic Coulomb Hamiltonian (1a) is also a cornerstone for wave mechanical solutions for chemical bonding. Unfortunately, the underlying Coulomb recipe is *hidden instead of made apparent* in approximations like (9c) of Pauling and Wilson for the molecular hydrogen cation and, by extension, of Heitler and London for molecular hydrogen, exactly as we argued in [1]. Simple mathematics as in (10c) suffices to undo this *unjust* hiding of fundamental process with 2 Coulomb quantum states ±1 (1a), (1f) in wave mechanics. We also showed explicitly why, to understand wave mechanics from a classical point of view, it is indeed helpful to replace the value of the wave function throughout by +1 [1]. Fig. 1 and Fig. 5 reveal that classical and wave mechanical bonding schemes are very similar indeed, although wave mechanics merely focuses on details (*distortions of the global scheme*) rather than on the global scheme itself. With the present *ab initio* analysis, it seems difficult to contest the reality of hydrogen-antihydrogen interactions even for explaining covalent bonding in molecular hydrogen [1]. With the results above, one should admit, sooner or later, that something very elementary went wrong indeed at a very early instance with *antimatter and its interactions with matter* [1].

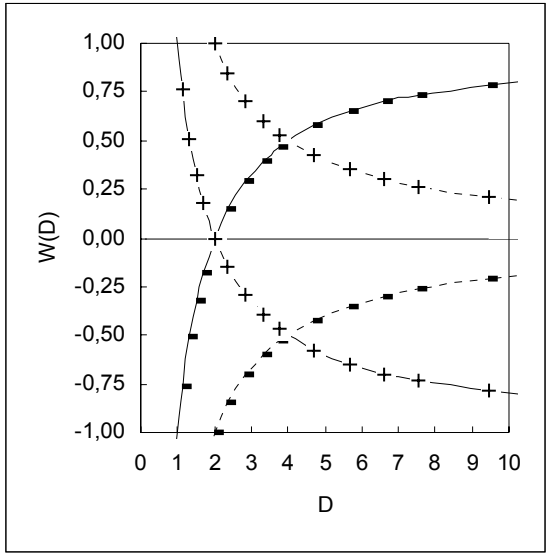

Fig. 1a Coulomb PECs with and without asymptotes versus D (- attractive, + repulsive)

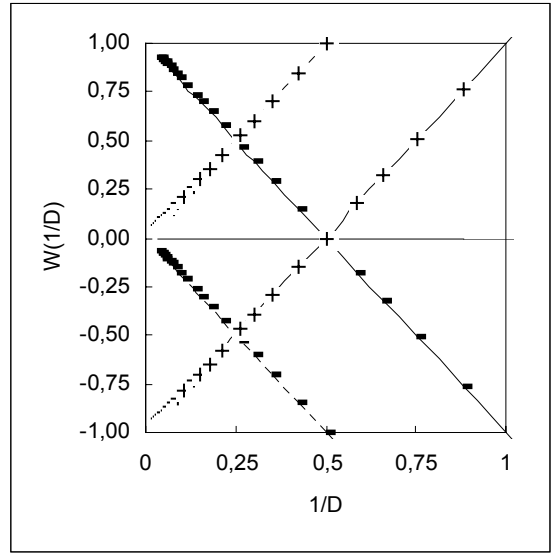

Fig. 1b Coulomb PECs with and without asymptotes versus 1/D (- attractive, + repulsive)

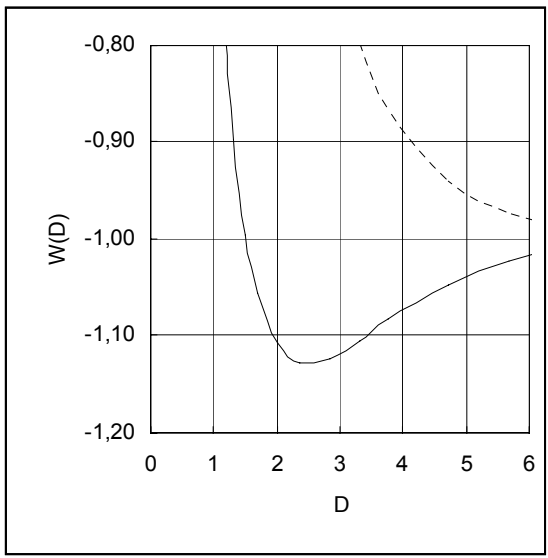

Fig. 2 PECs for $W_S$ and $W_A$ states of molecular hydrogen cation with wave mechanics (units $e^2/2a_0$) versus D (A slashes, S full)

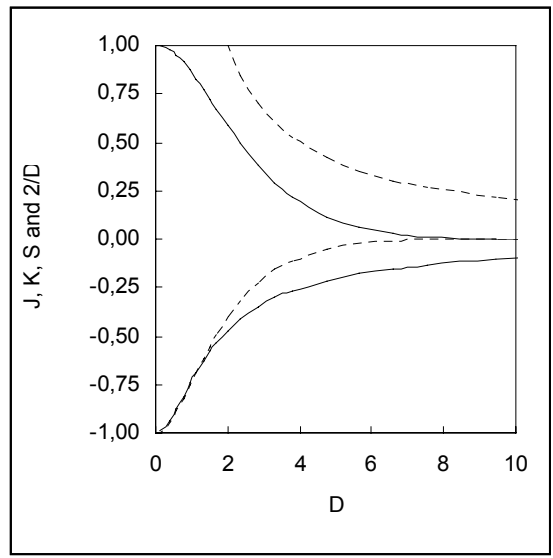

Fig. 3 Integrals J, K, S and 2/D versus D (attractive J full, K slashes; repulsive S full, 2/D slashes)

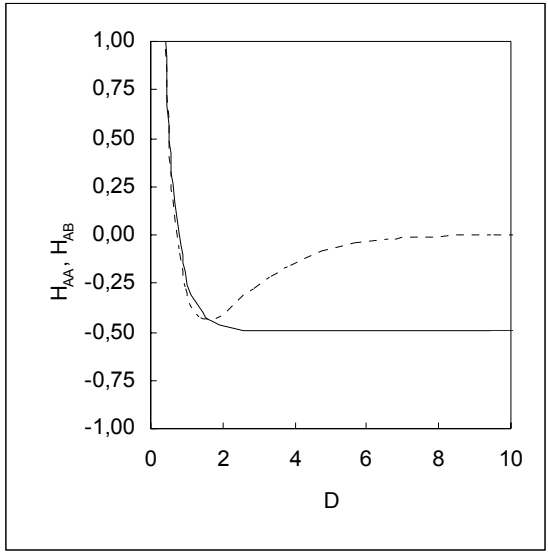

Fig. 4 $H_{AA}$ (full) and $H_{AB}$ (dashes) versus D

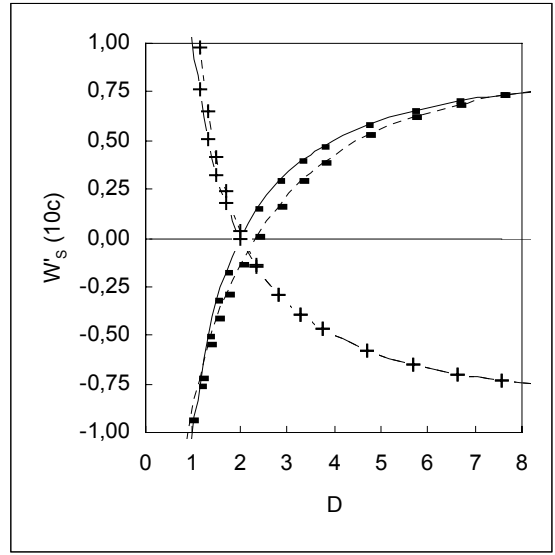

Fig. 5a $W_\pm$, eqn (1f) and $W'_S$, eqn (10c) versus D (+ repulsive, full 1f, dashes 10c; − attractive, full 1f, dashes 10c)

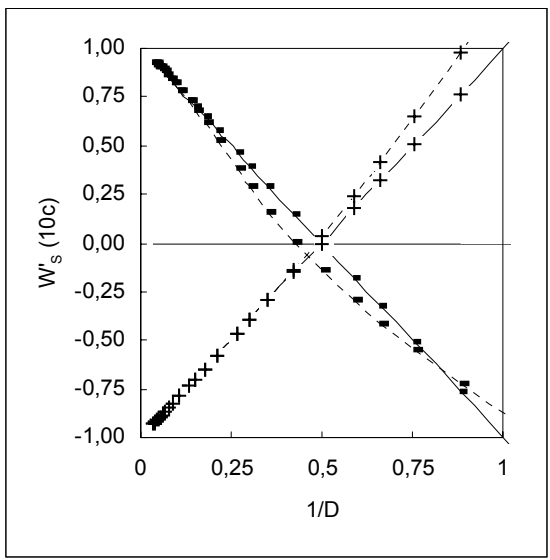

Fig. 5a $W_\pm$, eqn (1f) and $W'_S$, eqn (10c) versus D (+ repulsive, full 1f, dashes 10c; − attractive, full 1f, dashes 10c)

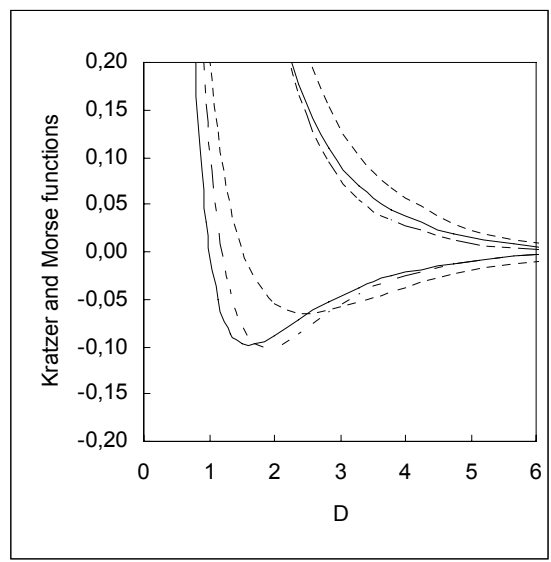

Fig. 6 Attenuated Kratzer (full lines); Morse (mixed dashes) and Pauling Wilson PECs (short dashes) versus D